\documentclass[a4paper]{spie}  
\usepackage[]{graphicx}
\usepackage{amstext} 
\usepackage{amssymb} 
\usepackage{amsmath}
\usepackage{txfonts}

\def\arcsec{\hbox{$^{\prime\prime}$}}

\newcommand*\arcmin{\ensuremath{^\prime}}
\newcommand*\sun{\ensuremath{\odot}}

\title{Astrometric detection of exoplanets from the ground}
\author{J.~Sahlmann\supit{a}, P.~F.~Lazorenko\supit{b}, A.~M\'erand\supit{c}, D. Queloz\supit{a}, D.~S\'egransan\supit{a}, J.~Woillez\supit{d}
\skiplinehalf
\supit{a}Observatoire de Gen\`eve, Universit\'e de Gen\`eve, 51 Chemin Des Maillettes, 1290 Versoix, Switzerland;\\
\supit{b}Main Astronomical Observatory, National Academy of Sciences of the Ukraine, Zabolotnogo 27, 03680 Kyiv, Ukraine;\\
\supit{c}European Southern Observatory, Alonso de C\'ordova 3107, Vitacura-Santiago, Chile;\\
\supit{d}European Southern Observatory, Karl-Schwarzschild-Str. 2, 85748 Garching, Germany}
\authorinfo{Further author information: (Send correspondence to J.S.)\\J.S.: E-mail: johannes.sahlmann@unige.ch}

\begin{document} 
  \maketitle 

\begin{abstract}
Astrometry is a powerful technique to study the populations of extrasolar planets around nearby stars. It gives access to a unique parameter space and is therefore required for obtaining a comprehensive picture of the properties, abundances, and architectures of exoplanetary systems. In this review, we discuss the scientific potential, present the available techniques and instruments, and highlight a few results of astrometric planet searches, with an emphasis on observations from the ground. In particular, we discuss astrometric observations with the Very Large Telescope (VLT) Interferometer and a programme employing optical imaging with a VLT camera, both aimed at the astrometric detection of exoplanets. Finally, we set these efforts into the context of Gaia, ESA's astrometry mission scheduled for launch in 2013, and present an outlook on the future of astrometric exoplanet detection from the ground. 
\end{abstract}

\section{INTRODUCTION}\label{sec:intro}
It is now established that extrasolar planets are abundant in the Galaxy \cite{Mayor:2011fj,Cassan:2012uq,Batalha:2013qf} and many hundreds have been discovered using a variety of techniques \cite{Seager:2011ve}. However, the degree to which exoplanets are characterised varies greatly. The goal of current large observational campaign is to obtain large samples of planets with uniformly and accurately determined properties. These are principally the orbital parameters, the planet's mass, size, and atmospheric characteristics, and the host star's properties. The comprehensive characterisation of planetary systems, which is necessary to progress in our understanding of planet formation and evolution, can generally only be obtained by combining several observing techniques.\\
Astrometry consists in measuring stellar positions in the sky and can be applied for indirect planet detection by determining the small lateral displacement of a star orbited by an unseen planet (measurements of relative astrometry are possible when the planet is detected directly, but those cases are not covered here). This contribution deals with the realisation of such observations from the ground, which have to cope with the disturbing influence of the Earth's atmosphere. We discuss the scientific motivation (Sect.~\ref{sec:pot}), introduce the employed techniques (Sect.~\ref{sec:tec}), present two ongoing projects using optical interferometry (Sect.~\ref{sec:pri}) and imaging (Sect.~\ref{sec:pal}), and conclude with an outlook into the future of ground-based astrometric planet searches (Sect.~\ref{sec:out}).     
	
\section{SCIENTIFIC POTENTIAL OF ASTROMETRY}\label{sec:pot}
The potential of astrometry for exoplanet detection was recognised when the available measurement precision reached a few milli-arcseconds (mas), thus offering the perspective of detecting giant Jupiter-like planets in wide orbits around nearby stars\cite{Gatewood:1976fk, Black:1982kx}. Unfortunately, most of the announced detections were later shown to be false and have taught us to be particularly cautious in this field. However, continued progress in instrumentation made first robust applications of astrometry possible and we refer the reader to the reviews in Refs. \citenum{Sozzetti:2005qy, Perryman:2012uq, Sahlmann:2012fk2}.\\ 
To illustrate the parameter space accessible with astrometric measurements, it is useful to point out some figures and attributes related to astrometry and to compare it with other planet detection techniques. For instance, a 10 Jupiter mass ($M_J$) companion in a 100 day orbit around a $1\,M_\sun$ star located at a distance of 10 pc induces an astrometric signature of the star's barycentric orbit of 0.4~mas, In comparison, the Hipparcos space mission\cite{Perryman:1989kx}, which is the only large survey of precision astrometry to date, realised a single measurement precision of $\sim$1~mas, thus would have difficulties to detect the 0.4 mas signal of our hypothetical planet. Formally, the angular semimajor axis $a_1$ of a star's barycentric orbit caused by a companion with mass $m_2$ depends on the distance to the system $d$, the orbital period $P$, and the primary mass $m_1$\cite{Hilditch:2001kx}
\begin{equation}\label{eq:1}
a_1 \propto \frac{m_2}{d} \left( \frac{P}{m_1+m_2} \right)^{{2}/{3}}.
\end{equation}
The amplitude of $a_1$, which is the signal we seek to detect, therefore decreases with increasing distance and primary mass
\begin{figure}
\begin{center} 
\includegraphics[width= 0.9\linewidth]{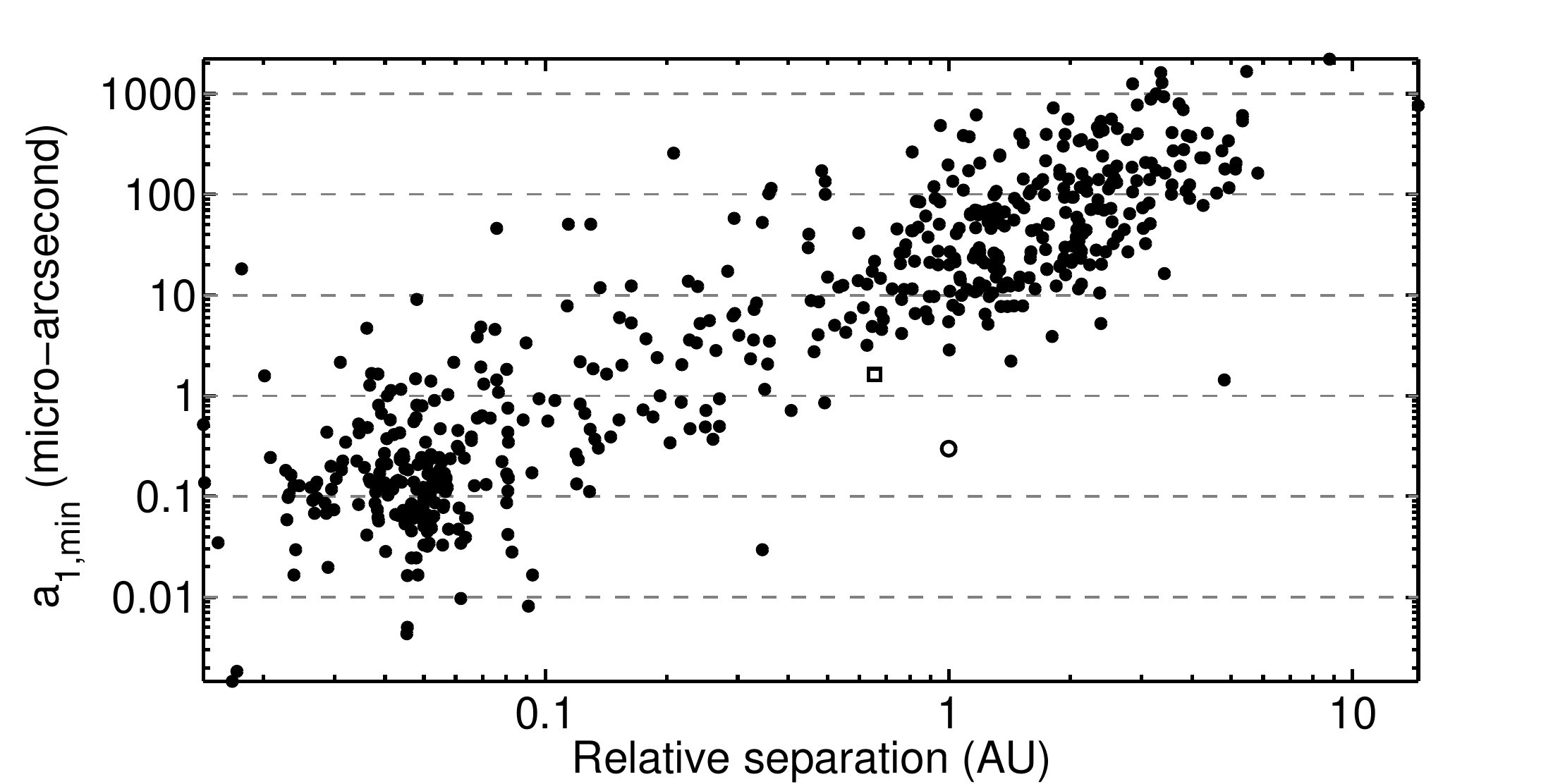}
 \end{center} 
\caption{Minimum astrometric signature for 570 known exoplanets as a function of relative separation from the host star (assuming circular orbits). The open circle marks the astrometric jitter of the Sun due to Earth when seen at a distance of 10 pc, which has a semi-amplitude of 0.3 $\mu$as. The open square shows the 1.6 $\mu$as signature of a hypothetical Earth-mass planet in the habitable zone of Alpha Centauri B, a planet hosting star\cite{Dumusque:2012fk} located at 1.3 pc.} \label{fig:a1min}
\end{figure}
 but increases for increasing orbital period and planet mass. This is different for the radial velocity (RV) and transit photometry techniques, whose detection capabilities are nearly independent of distance. However, the RV signal of a given planet decreases with orbital period and, assuming that planetary orbits are randomly oriented in space, the transit probability also decreases with the period. Direct imaging and gravitational microlensing searches were so far predominantly sensitive to wide orbits, and usually do not well constrain all orbital parameters. The relation (\ref{eq:1}) shows that  the measurement of $a_1$ determines the planet mass, provided that the primary mass can be estimated. Astrometry resolves the 2-dimensional orbit and therefore gives us access to all seven orbital parameters and thus to a planet-mass estimate without the $\sin i$ ambiguity present in RV planet masses.\\
It is also instructive to inspect the astrometric signatures of the known population of exoplanets. Figure~\ref{fig:a1min} shows the minimum semimajor axes of the barycentric stellar orbits caused by 570 planets listed in the exoplanet.org database\cite{Wright:2011lr} on July 4, 2013, that have an entry for planet mass, stellar mass, distance, and orbital period. These are lower limits to $a_1$, because we assumed $m_2 = m_2\,\sin i$ when the inclination was unknown, which is the case for most RV planets. The vertical scatter stems from different distances and primary/secondary masses of systems with the same relative separation. Keeping in mind that the planet compilation is subject to biases, we observe that only five planets (i.e. $<\,$1~\%) introduce an astrometric signal larger than 1~mas. Lowering the threshold to 100 and 10 micro-arcseconds ($\mu$as) increases the number of known planets with signatures greater than that to 101 and 271, respectively.\\ 
We now can broadly summarise the requirements and the potential of astrometry with the following qualitative statements.
\begin{enumerate}
  \item To make a significant contribution to the research field of extrasolar planets by determining the orbits of host stars, an astrometric study typically has to reach an accuracy of $\sim$10-100 $\mu$as.
  \item Provided that an accuracy well below 1 mas is achieved, astrometric surveys can offer invaluable information on planetary systems and open potentially unique detection opportunities. Astrometry is particularly suited to accurately derive the planet mass function (without $\sin i$ ambiguity) and to study long-period planets ($P$$\sim$1-10 years, taking advantage of larger amplitudes). In addition, some parameters of planetary systems are nearly exclusively accessible by astrometry, e.g. the mutual orbital inclinations in non-transiting multi-planet systems\cite{McArthur:2010kx} and spin-orbit obliquities for non-transiting planets\cite{Sahlmann:2011lr}.
  \item The astrometric accuracy required to discover and study planets similar to Earth around nearby stars, which is one of the outstanding goals of astronomy, is $\sim$0.1-1 $\mu$as (see Fig.~\ref{fig:a1min}), i.e. 2-3 orders of magnitude better than what currently is available. However, astrometry is a promising technique to pursue this goal because it is expected to be less prone to perturbations by stellar activity compared to other techniques\cite{Eriksson:2007uq} and, due to the complete orbit characterisation, it makes efficient follow-up observations possible\cite{Malbet:2011fk}.
\end{enumerate}

\section{GROUND-BASED TECHNIQUES}\label{sec:tec}
The position of stars can be measured either in an absolute sense by directly relating them to a global reference frame or relative to a local coordinate system, usually constituted by other stars in the field of view. In either case, astrometric observations from the Earth's surface are affected by turbulence occurring in the atmosphere that alters the instantaneous apparent position of a star. Second, systematic errors of the measurement process have to be mitigated carefully, which calls for detailed calibration and operation procedures. Finally and fundamentally, the information content of the observations, often measured in terms of signal-to-noise (S/N), has to be sufficient for the targeted measurement precision $\delta$ 
\begin{equation}
\delta \propto \frac{1}{\mathrm{S/N}}\frac{\lambda}{D},
\end{equation}
where $\lambda$ is the observing wavelength and $D$ is the aperture size. In practice, the control of systematic errors is the limiting factor of modern ground-based astrometry. Note that we will limit the discussion mostly to optical techniques (which includes the near-infrared spectral range) and only give a brief overview of radio VLBI applications in Sect.~\ref{sec:vlbi}.

\subsection{The limitations imposed by atmospheric turbulence}
The refractive index of the air column along the line of sight of an astronomical observatory is variable mostly due to temperature fluctuations\cite{Roddier:1999ly}. An incoming wavefront of stellar light therefore experiences phase distortions across the telescope pupil, which affect the location of the star's image formed in the focal plane of the instrument. Wind alters the instantaneous phase distortions with the effect of random image motion, which depends on the spatial and temporal variation of the refractive index.\\
Because the atmospheric turbulence is entirely uncorrelated over wide fields, global optical astrometry from the ground is limited to $\sim$10 mas\cite{Shao:1990qq}, thus inhibiting exoplanet science. A much better performance can be achieved in narrow-angle mode by measuring the relative positions of stellar objects within a small field of view, typically $\lesssim$1\arcmin~for optical observations. In this narrow-angle mode, the variance $\delta^2$ of the measured angular separation $\Theta$ depends on the instrument aperture size $D$, and the exposure time $T$
\begin{equation}\label{eq:2} 
\delta^2 \propto {\Theta^r}\left({D^p T}\right)^{-1}. 
\end{equation}
In line with classic Kolmogorov turbulence, the commonly used parameter values are $r=2$ and $p=4/3$\cite{Lindegren:1980bu,Han:1989ys}, which also apply to optical interferometry\cite{Shao1992} and correspond to the case of a single reference object. However, with several reference stars, the correlated turbulence within a field of size $\Theta$ can be suppressed much stronger. This requires symmetrisation of the reference star field, after which the model parameters were shown to be $r = 11/3$ and $p = 3$, thus leading to a much improved performance for large filled apertures of size $D$\cite{Lazorenko:2002qy,Lazorenko:2002lr,Lazorenko:2004cs,Lazorenko:2009ph}. These calculations predict that both optical long baseline interferometry and imaging with an 8 m-class telescope are capable of realising 10-100 $\mu$as astrometry with $\sim$1/2 hour-long exposures in narrow-angle mode.

\subsection{Imaging astrometry}
The most basic astrometric technique is to take an image of a group of stars and to measure their relative positions. The goal is to measure the proper, parallactic, and orbital motion of a nearby star with a planet relative to the surrounding stars located at large distances. Like in other fields of astronomy, the use of CCDs made steep improvements in performance and efficiency possible\cite{Monet:1983vn}. Today, both seeing-limited and adaptive-optics assisted cameras are being used for exoplanet work. The Palomar 5 m and the 2.5~m Du Pont telescope are used for astrometric planet searches around late-type stars\cite{Pravdo:1996fk, Boss:2009ff}. With precisions of $\sim$0.5-2 mas these programmes could detect brown dwarf companions of stars\cite{Pravdo:2005fu} and constrain the mass range of the planet around GJ\,317\cite{Anglada-Escude:2012vn}. The best performance for seeing limited observations is currently achieved with the Very Large Telescope (VLT), which reaches an accuracy of $\lesssim$ 200 $\mu$as and is discussed in Sect.~\ref{sec:pal}.\\
Near-infrared cameras equipped with adaptive optics systems can reach astrometric precisions of 100-300 $\mu$as\cite{Neuhauser:2007lr,Cameron:2009eu, Fritz:2010lr, Kervella:2013uq}. However, planet search surveys using these instruments were so far limited to arcsecond-scale binaries (using one binary component as astrometric reference and searching for orbital motion of the other component) and have not yet yielded a detection\cite{Roll:2011fk}.

\subsection{Astrometry with optical interferometry}
The large effective aperture sizes that can be generated with the multiple telescopes of an optical interferometer make them ideal instruments for high-precision astrometry. Usually, only two stars can be observed jointly with this technique and one serves as astrometric reference for searching the orbital motion of the other. The prospect of reaching 10 $\mu$as precision for relative astrometry of two stars\cite{Shao1992} separated by $\lesssim1\arcmin$ and the attractive science case of exoplanet detection led to the successful demonstration experiment at the Palomar Testbed Interferometer \cite{Lane2000}(PTI). The Keck and VLT interferometer implemented this mode of operation with the ASTRA\cite{Woillez:2010rt} and PRIMA\cite{Sahlmann:2013fk} facilities, respectively. Because both PTI and the Keck Interferometer have ceased operation, PRIMA is the only facility to currently implement this mode and we discuss it in more detail in Sect.~\ref{sec:pri}.\\
A special observing mode (the `very narrow-angle` mode) applicable to the planet search around stars in sub-arcsecond binaries consists in scanning the two fringe packets resolved by the interferometer in delay space to determine the stars' relative separation. Precisions of 10 $\mu$as were achieved\cite{Lane:2004rm} and yielded the currently best candidate for an exoplanet detected by astrometry alone: a $\sim$2 Jupiter-mass planet candidate in a 1000 day orbit around one of the components of HD\,176051\cite{Muterspaugh:2010lr2}. A project employing a similar strategy is currently underway at the Sydney University Stellar Interferometer\cite{Kok:2013uq}.

\subsection{Radio Very Long Baseline Interferometry (VLBI)}\label{sec:vlbi}
Radio VLBI observations in phase-referencing mode (analogue to narrow-angle interferometry in the optical) routinely yield relative astrometry at the 10-100 $\mu$as accuracy level\cite{Fomalont:2003ys, Reid:2004ve, Reid:2009uq}. Unfortunately for the prospect of exoplanet detection, most nearby stars are not bright enough at radio wavelength to be studied with this technique. However, planet search efforts have been reported for M dwarfs\cite{Bower:2011fj, Forbrich:2013uq} and promote VLBI astrometry as a promising technique for specific questions in exoplanet science.

\section{ESPRI: ASTROMETRIC SEARCH FOR PLANETS WITH PRIMA}\label{sec:pri}
The goal of the ESPRI project\cite{Launhardt2008} is to discover and characterise extrasolar planets around nearby stars using astrometric observations with the VLT interferometer (VLTI). The dual-feed infrastructure that makes such observations possible at the VLTI is named PRIMA. The ESPRI survey was designed to last several years and to address three main topics: the determination of orbital inclinations for nearby RV planets, the search for long-period giant planets in the solar neighbourhood, and the search for giant planets around young stars\cite{Launhardt2008}.

\subsection{Principles}
An exhaustive description of the principles, the instrument hardware, and the operation of the PRIMA astrometric instrument is given in Ref. \citenum{Sahlmann:2013fk}. Additional background information can be found in Refs. \citenum{Sahlmann:2012fk2,Sahlmann:2012uq} and we only give a basic overview here: PRIMA implements the dual-feed capability at the VLTI and makes it possible to observe two stars separated by $\lesssim1\arcsec$ simultaneously with two 1.8 m Auxiliary Telescopes spanning a $\sim$100 m baseline. Long-stroke delay lines and short-stroke differential delay lines compensate the optical delays internally and the stellar beams are combined in the two fringe sensors (one per star), who drive two feedback loops for fringe tracking. A 4-beam laser metrology measures the effective differential delay which is used to determine the projected angular separation of the two stars.  

\subsection{Status}
\begin{figure}[h]
\begin{center} 
\includegraphics[width= 0.5\linewidth]{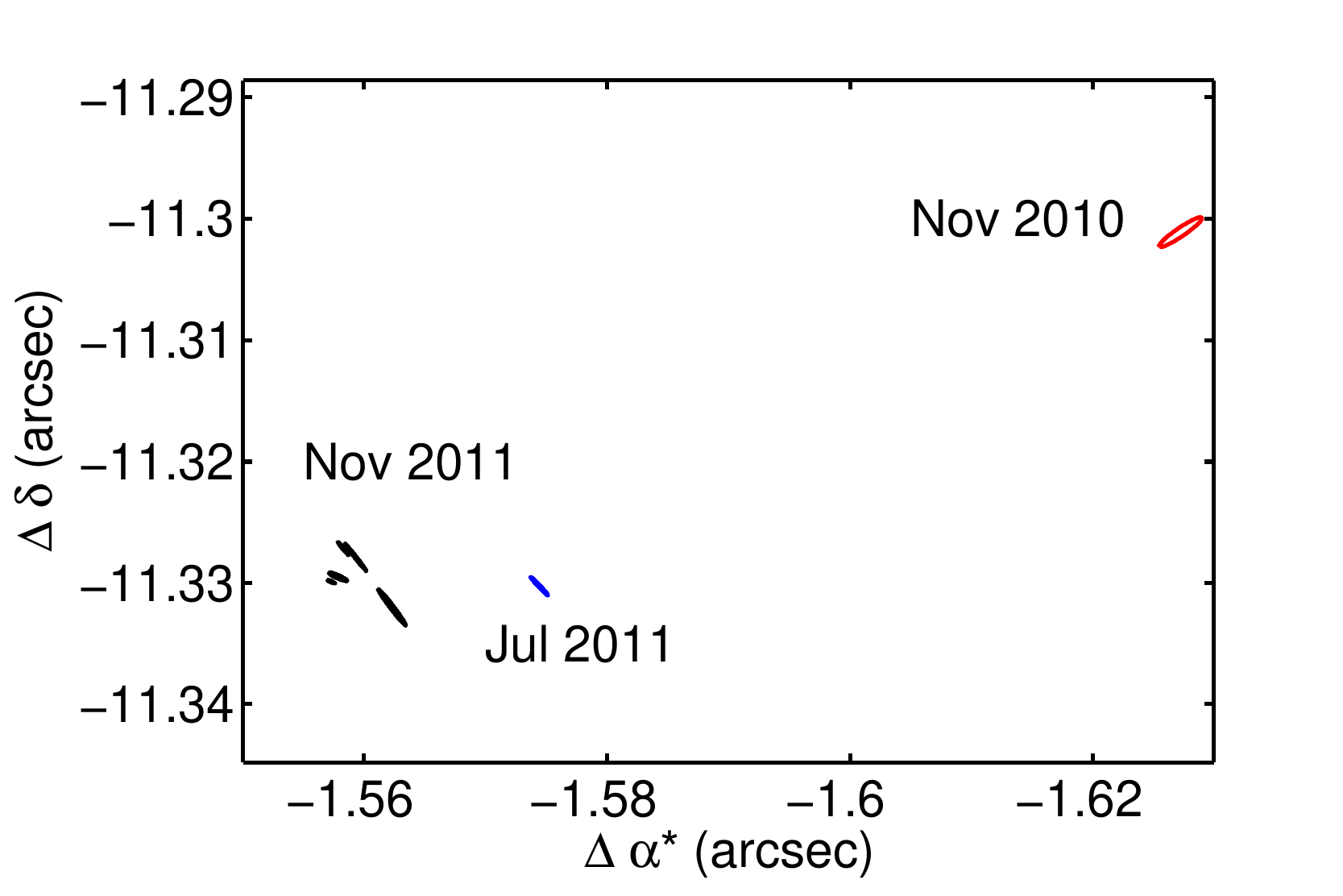}
 \end{center} 
\caption{The orbital motion of the visual binary HD\,10360 as measured by PRIMA before the metrology upgrade. North is up and East is left. The measured positions of HD\,10361 relative to HD\,10360 (located 11\arcsec~North) are shown with 3-$\sigma$ error ellipses for three measurement campaigns over one year. The (uncertain) orbital motion is followed qualitatively but the Nov. 2011 measurements reveal a mas-scale scatter. This figure was adapted from Ref. \citenum{Sahlmann:2013fk}.}
\label{fig:prima}
\end{figure}

In 2011, PRIMA was used to perform astrometric test observations of bright visual binaries aimed at exercising the operational procedures and establishing the basic performances (Fig.~\ref{fig:prima}). For target separations $<$10\arcsec and on timescales $<$1 h, the astrometric precision of PRIMA was shown to be limited predominantly by the atmosphere to a level of 20-30 $\mu$as. For wide-separation targets, however, much larger systematic errors were recorded and the night-to-night repeatability of the astrometric measurement was limited to a few mas\cite{Sahlmann:2013fk}. A start of scientific operations for the ESPRI survey was therefore unconceivable and the project moved to a phase of re-assessment and trouble shooting. The dominant source of systematic errors was identified as the beamtrain between the telescope primary mirror and the ninth mirror in order of incidence (M9) that previously was not monitored by the laser metrology system. Consequently, the alignment of some VLTI/PRIMA subsystems was improved and the laser beampath was extended up to the secondary telescope mirror, which required substantial hardware changes. At the time of writing (July 2013), the so modified system is being tested on-sky. 

\section{PRECISION ASTROMETRY WITH FORS2 AT THE VLT}\label{sec:pal}
In a series of papers since 2006, P. Lazorenko and collaborators have shown that a specially tailored reduction method applied to images taken with the FORS2\cite{Appenzeller:1998lr} camera mounted on one of the 8 m telescopes of the VLT can lead to astrometric measurements of excellent precision and stability\cite{Lazorenko:2006qf,Lazorenko:2007ul,Lazorenko:2009ph}. The demonstrated precision reaches 50 $\mu$as in some cases, which makes exoplanet search and characterisation possible with this instrument. Below, we briefly review the principles of the technique and discuss the results published so far.

\subsection{Principles}
The main requirements to achieve $\sim$100 $\mu$as astrometry with seeing-limited images of a single-dish optical telescope can be summarised as follows\cite{Lazorenko:2002qy,Lazorenko:2002lr,Lazorenko:2004cs,Lazorenko:2006qf,Lazorenko:2007ul,Lazorenko:2009ph}: (1) An 8 m-class telescope is required to average out the atmospheric image motion. (2) A large number ($\gtrsim100$) of reference stars with magnitudes similar to the science target have to be present in the field-of-view (typically a few arc-minutes wide). (3) Spatial distortions of the image point-spread-function have to be modeled carefully and their variation in time has to be minimised by restricting the allowable seeing conditions. (4) Image motions and distortions by the atmosphere and the instrument/telescope have to be mitigated on the basis of the reference star field. By combining all those requirements, a long-term accuracy of 50 $\mu$as has been demonstrated\cite{Lazorenko:2009ph}.

\subsection{Results}
\begin{figure}[h]
\begin{center} 
\includegraphics[width= 0.9\linewidth]{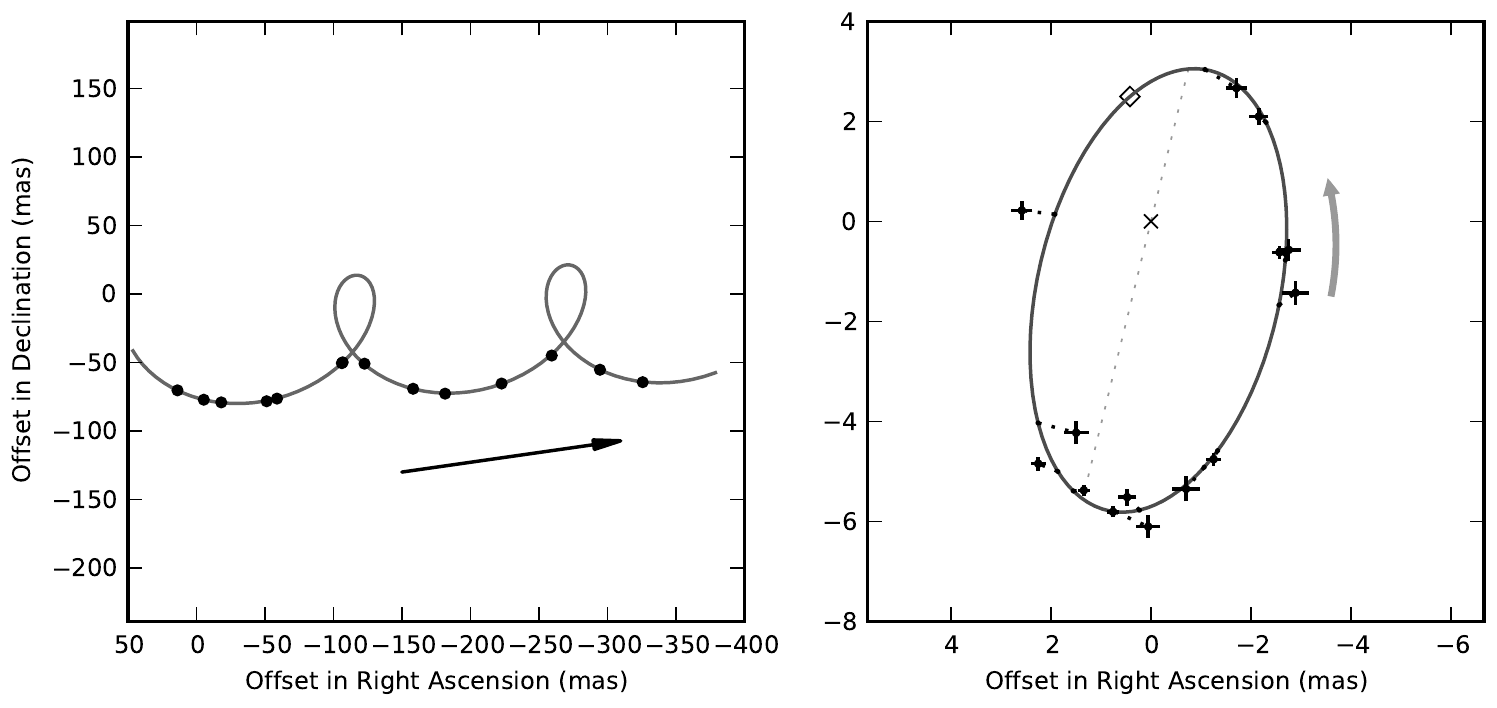}
 \end{center} 
\caption{\emph{Left}: Proper and parallactic motion of DE0823--49 measured with FORS2. Black circles indicate the 14 measurements and the grey curve shows the best-fit model. \emph{Right}: The barycentric orbit of DE0823--49 caused by a 28 $M_{Jup}$ companion. Measurements are shown with 1-$\sigma$ errorbars, which illustrates the high S/N of this detection. These figures were adapted from Ref. \citenum{Sahlmann:2013kk}.}
\label{fig:orbit}
\end{figure}

For the planet searcher, the requirement (2) above turns out to be the most constraining because it limits the type of attainable targets. Very low-mass stars and brown dwarfs are ideal targets, because the are nearby, thus promising large angular reflex motions, but also faint so that sufficient background stars are usable in the images. Even more importantly, the question of planet occurrence around these objects is unresolved.\\ 
Those 'ultracool' dwarfs are therefore the first targets to be studied with FORS2 and the initial astrophysical application was the independent disproof of the giant planet around VB\,10 using only four astrometric measurements with an average precision of 90 $\mu$as collected over $\sim$17 days\cite{Lazorenko:2011lr}. At about the same time, an astrometric planet search survey targeting nearby ultracool dwarfs was initiated. The true potential of FORS2 astrometry was demonstrated with the first result of this survey, the discovery of a brown dwarf companion in a 246 day orbit around the L1.5 dwarf DE0823--49\cite{Sahlmann:2013kk} (Fig.~\ref{fig:orbit}). The precision and accuracy of the method was therewith shown to be sufficient to detect potential planets of a few Neptune masses in $>$500 day orbits around nearby ultracool dwarfs.

\section{FUTURE PROSPECTS}\label{sec:out}
Astrometric planet searches from the ground have so far yielded few but promising results, e.g. a good candidate for the first astrometric planet detection\cite{Muterspaugh:2010lr2} and an unusual very low-mass binary system\cite{Sahlmann:2013kk}. If the ESPRI project succeeds in overcoming the instrument-related problems and PRIMA astrometry will be limited solely by the atmosphere, we can expect results on giant planets in the near future. In addition, new instruments and techniques\cite{Guyon:2012fk} are emerging and provide us with the astrometric accuracy necessary for planet detection. One instrument that will certainly boost this field of research is the Gaia space astrometry mission. 

\subsection{Impact of Gaia}
Gaia is a space mission and therefore slightly out of the scope of this review, but it will revolutionise the application of astrometry to exoplanet science, thus has to be discussed here. Gaia\cite{Perryman:2001vn, de-Bruijne:2012kx} is a cornerstone mission of the European Space Agency and is expected to be launched in November 2013. Following observation principles similar to its predecessor Hipparcos, Gaia will create an all-sky map of about one billion stellar objects with visible magnitudes of $\sim$6-20. On average, every object will be observed $\sim$70 times during the 5-year mission lifetime, yielding a wealth of astrometric, photometric, and spectroscopic data. The expected single-measurement astrometry precision of $\sim$20-50 $\mu$as in the $G$$\sim$6-15 magnitude range\cite{2011AdSpR..47..356M} makes Gaia an exoplanet discovery machine with an expected yield of several thousand new giant planets with periods $\lesssim$10 years around stars within $\lesssim$\,200 pc\cite{Casertano:2008th}.\\ 
We can therefore expect that Gaia will make astrometry a standard tool for the study of exoplanetary systems. Its uniform survey and large number of detected planets will make statistical studies of planet and host star parameters possible and thereby lead to new insights into the planet formation and evolution processes.

\subsection{Future potential of ground-based astrometry}
Although Gaia will probably dominate this field over the next few years, ground-based efforts should by no means be cut back. On the contrary, we need efficient ground-based facilities to independently address numerous science questions and for the preparation and follow-up of space missions that have a limited lifetime. Ground-based facilities have been and will remain essential for the development and demonstration of new methodologies and technologies.\\
The space and ground-based optical imaging telescopes differ in the size of their entrance aperture and consequently in their light collecting power. Therefore, whereas astrometry of bright objects is advantageous from space, the monitoring of faint stars can be favourable with large ground-based telescopes. This is illustrated by FORS2/VLT observations that have a single-epoch precision comparable to the Gaia expectation on stars fainter than $\sim$15th magnitude\cite{Sahlmann:2013kk, 2011AdSpR..47..356M}. This comparison testifies for a good performance of the currently working  seeing-limited imaging telescopes for the planet search around faint nearby stars.\\
The astrometric reduction used for the FORS2 observations discussed above effectively eliminates atmospheric image motion and is limited primarily by the seeing variations and the S/N of the star images. Using adaptive optics systems with wide-field correction as those coming online on 8 m-class telescopes\cite{Meyer:2011fj, Rigaut:2012zr, Paufique:2012kx}, the S/N can be increased and we expect that the precision of dedicated observations can reach $\sim$10\,$\mu$as. Similarly, the 30-40 m apertures of future extremely large telescopes (e.g.\ E-ELT, TMT) offer the opportunity for extremely precise astrometry, potentially even better than 10 $\mu$as\cite{Lazorenko:2004cs,Trippe:2010fj}, provided that their cameras cover a field wide enough to capture a large number of reference stars.\\
Because of the precision requirement of better than 0.5 $\mu$as, it is unlikely that ground-based astrometry will be capable to detect Earth twins in the near future and dedicated space mission will probably be necessary\cite{Unwin:2008ys, Malbet:2011fk}. However, we should keep in mind that the present-day performance of $<$1 m/s of stellar radial velocity measurements was equally unimaginable some decades ago.

\acknowledgments
J.S., D.Q., and D.S. thank the Swiss National Science Foundation for supporting this research. J.S. kindly acknowledges support as a visitor at the Centro de Astrobiolog\'ia in Villanueva de la Ca\~nada (Madrid).     

\bibliography{spie2013}   

\begin{thebibliography}{10}

\bibitem{Mayor:2011fj}
{Mayor}, M., {Marmier}, M., {Lovis}, C., et~al., ``{The HARPS search for
  southern extra-solar planets XXXIV. Occurrence, mass distribution and orbital
  properties of super-Earths and Neptune-mass planets},'' {\em ArXiv e-prints}
  (2011).

\bibitem{Cassan:2012uq}
{Cassan}, A., {Kubas}, D., {Beaulieu}, J.-P., et~al., ``{One or more bound
  planets per Milky Way star from microlensing observations},'' {\em \nat}~{\bf
  481},  167--169 (2012).

\bibitem{Batalha:2013qf}
{Batalha}, N.~M., {Rowe}, J.~F., {Bryson}, S.~T., et~al., ``{Planetary
  Candidates Observed by Kepler. III. Analysis of the First 16 Months of
  Data},'' {\em \apjs}~{\bf 204},  24 (2013).

\bibitem{Seager:2011ve}
{Seager, S.}, ed.,  [{\em {Exoplanets}}{\nolinebreak\hspace{0.1em}]},
  University of Arizona Press (2011).

\bibitem{Gatewood:1976fk}
{Gatewood}, G., ``{On the astrometric detection of neighboring planetary
  systems},'' {\em \icarus}~{\bf 27},  1--12 (1976).

\bibitem{Black:1982kx}
{Black}, D.~C. and {Scargle}, J.~D., ``{On the detection of other planetary
  systems by astrometric techniques},'' {\em \apj}~{\bf 263},  854--869 (1982).

\bibitem{Sozzetti:2005qy}
{Sozzetti}, A., ``{Astrometric Methods and Instrumentation to Identify and
  Characterize Extrasolar Planets: A Review},'' {\em \pasp}~{\bf 117},
  1021--1048 (2005).

\bibitem{Perryman:2012uq}
{Perryman}, M., ``{The history of astrometry},'' {\em European Physical Journal
  H}~{\bf 37},  745--792 (2012).

\bibitem{Sahlmann:2012fk2}
{Sahlmann}, J., {\em {Observing exoplanet populations with high-precision
  astrometry}}, PhD thesis, Observatoire de Gen{\`e}ve, Universit{\'e} de
  Gen{\`e}ve (2012).

\bibitem{Perryman:1989kx}
{Perryman}, M.~A.~C., ``{Hipparcos: astrometry from space},'' {\em \nat}~{\bf
  340},  111--116 (1989).

\bibitem{Hilditch:2001kx}
{Hilditch}, R.~W.,  [{\em {An Introduction to Close Binary
  Stars}}{\nolinebreak\hspace{0.1em}]}, Cambridge University Press (2001).

\bibitem{Dumusque:2012fk}
{Dumusque}, X., {Pepe}, F., {Lovis}, C., et~al., ``{An Earth-mass planet
  orbiting $\alpha$ Centauri B.},'' {\em \nat}~{\bf 491} (2012).

\bibitem{Wright:2011lr}
{Wright}, J.~T., {Fakhouri}, O., {Marcy}, G.~W., et~al., ``{The Exoplanet Orbit
  Database},'' {\em \pasp}~{\bf 123},  412--422 (2011).

\bibitem{McArthur:2010kx}
{McArthur}, B.~E., {Fritz.~Benedict}, G., {Barnes}, R., et~al., ``{New
  Observational Constraints on the {$\upsilon$} Andromedae System with Data
  from the Hubble Space Telescope and Hobby-Eberly Telescope},'' {\em
  \apj}~{\bf 715},  1203--1220 (2010).

\bibitem{Sahlmann:2011lr}
{Sahlmann}, J., {Lovis}, C., {Queloz}, D., et~al., ``{HD 5388 b is a 69
  $M_{Jup}$ companion instead of a planet},'' {\em \aap}~{\bf 528} (2011).

\bibitem{Eriksson:2007uq}
{Eriksson}, U. and {Lindegren}, L., ``{Limits of ultra-high-precision optical
  astrometry. Stellar surface structures},'' {\em \aap}~{\bf 476},  1389--1400
  (2007).

\bibitem{Malbet:2011fk}
{Malbet}, F., {L\'eger}, A., {Shao}, M., et~al., ``{High precision astrometry
  mission for the detection and characterization of nearby habitable planetary
  systems with the Nearby Earth Astrometric Telescope (NEAT)},'' {\em
  Experimental Astronomy} ,  109 (2011).

\bibitem{Roddier:1999ly}
{Roddier}, F.,  [{\em {Adaptive optics in
  astronomy}}{\nolinebreak\hspace{0.1em}]}, Cambridge University Press (1999).

\bibitem{Shao:1990qq}
{Shao}, M., {Colavita}, M.~M., {Hines}, B.~E., et~al., ``{Wide-angle astrometry
  with the Mark III stellar interferometer},'' {\em \aj}~{\bf 100},  1701--1711
  (1990).

\bibitem{Lindegren:1980bu}
{Lindegren}, L., ``{Atmospheric limitations of narrow-field optical
  astrometry},'' {\em \aap}~{\bf 89},  41--47 (1980).

\bibitem{Han:1989ys}
{Han}, I., ``{The accuracy of differential astrometry limited by the
  atmospheric turbulence},'' {\em \aj}~{\bf 97},  607--610 (1989).

\bibitem{Shao1992}
{Shao}, M. and {Colavita}, M.~M., ``{Potential of long-baseline infrared
  interferometry for narrow-angle astrometry},'' {\em \aap}~{\bf 262},
  353--358 (1992).

\bibitem{Lazorenko:2002qy}
{Lazorenko}, P.~F., ``{Differential image motion at non-Kolmogorov distortions
  of the turbulent wave-front},'' {\em \aap}~{\bf 382},  1125--1137 (2002).

\bibitem{Lazorenko:2002lr}
{Lazorenko}, P.~F., ``{Non-Kolmogorov features of differential image motion
  restored from the Multichannel Astrometric Photometer data},'' {\em
  \aap}~{\bf 396},  353--360 (2002).

\bibitem{Lazorenko:2004cs}
{Lazorenko}, P.~F. and {Lazorenko}, G.~A., ``{Filtration of atmospheric noise
  in narrow-field astrometry with very large telescopes},'' {\em \aap}~{\bf
  427},  1127--1143 (2004).

\bibitem{Lazorenko:2009ph}
{Lazorenko}, P.~F., {Mayor}, M., {Dominik}, M., et~al., ``{Precision
  multi-epoch astrometry with VLT cameras FORS1/2},'' {\em \aap}~{\bf 505},
  903--918 (2009).

\bibitem{Monet:1983vn}
{Monet}, D.~G. and {Dahn}, C.~C., ``{CCD astrometry. I - Preliminary results
  from the KPNO 4-m/CCD parallax program},'' {\em \aj}~{\bf 88},  1489--1507
  (1983).

\bibitem{Pravdo:1996fk}
{Pravdo}, S.~H. and {Shaklan}, S.~B., ``{Astrometric Detection of Extrasolar
  Planets: Results of a Feasibility Study with the Palomar 5 Meter
  Telescope},'' {\em \apj}~{\bf 465},  264--+ (1996).

\bibitem{Boss:2009ff}
{Boss}, A.~P., {Weinberger}, A.~J., {Anglada-Escud{\'e}}, G., et~al., ``{The
  Carnegie Astrometric Planet Search Program},'' {\em \pasp}~{\bf 121},
  1218--1231 (2009).

\bibitem{Pravdo:2005fu}
{Pravdo}, S.~H., {Shaklan}, S.~B., and {Lloyd}, J., ``{Astrometric Discovery of
  GJ 802b: In the Brown Dwarf Oasis?},'' {\em \apj}~{\bf 630},  528--534
  (2005).

\bibitem{Anglada-Escude:2012vn}
{Anglada-Escud{\'e}}, G., {Boss}, A.~P., {Weinberger}, A.~J., et~al.,
  ``{Astrometry and Radial Velocities of the Planet Host M Dwarf GJ 317: New
  Trigonometric Distance, Metallicity, and Upper Limit to the Mass of GJ
  317b},'' {\em \apj}~{\bf 746},  37 (2012).

\bibitem{Neuhauser:2007lr}
{Neuh{\"a}user}, R., {Seifahrt}, A., {R{\"o}ll}, T., et~al., ``{Detectability
  of Planets in Wide Binaries by Ground-Based Relative Astrometry with AO},''
  in [{\em IAU Symposium}{\nolinebreak\hspace{0.1em}]},   {\bf 240},  261--263
  (2007).

\bibitem{Cameron:2009eu}
{Cameron}, P.~B., {Britton}, M.~C., and {Kulkarni}, S.~R., ``{Precision
  Astrometry With Adaptive Optics},'' {\em \aj}~{\bf 137},  83--93 (2009).

\bibitem{Fritz:2010lr}
{Fritz}, T., {Gillessen}, S., {Trippe}, S., et~al., ``{What is limiting
  near-infrared astrometry in the Galactic Centre?},'' {\em \mnras}~{\bf 401},
  1177--1188 (2010).

\bibitem{Kervella:2013uq}
{Kervella}, P., {M{\'e}rand}, A., {Petr-Gotzens}, M.~G., et~al., ``{The nearby
  eclipsing stellar system {$\delta$} Velorum. IV. Differential astrometry with
  VLT/NACO at the 100 microarcsecond level},'' {\em \aap}~{\bf 552},  A18
  (2013).

\bibitem{Roll:2011fk}
{R{\"o}ll}, T., {Seifahrt}, A., {Neuh{\"a}user}, R., et~al., ``{Ground based
  astrometric search for extrasolar planets in stellar multiple systems},'' in
  [{\em EAS Publications Series}{\nolinebreak\hspace{0.1em}]},   {\bf 45},
  429--432 (2011).

\bibitem{Lane2000}
{Lane}, B.~F., {Colavita}, M.~M., {Boden}, A.~F., et~al., ``{Palomar Testbed
  Interferometer: update},'' in [{\em Society of Photo-Optical Instrumentation
  Engineers (SPIE) Conference Series}{\nolinebreak\hspace{0.1em}]},   {\bf
  4006} (2000).

\bibitem{Woillez:2010rt}
{Woillez}, J., {Akeson}, R., {Colavita}, M., et~al., ``{ASTRA: astrometry and
  phase-referencing astronomy on the Keck interferometer},'' in [{\em Proc.
  SPIE}{\nolinebreak\hspace{0.1em}]},   {\bf 7734} (2010).

\bibitem{Sahlmann:2013fk}
{Sahlmann}, J., {Henning}, T., {Queloz}, D., et~al., ``{The ESPRI project:
  astrometric exoplanet search with PRIMA. I. Instrument description and
  performance of first light observations},'' {\em \aap}~{\bf 551},  A52
  (2013).

\bibitem{Lane:2004rm}
{Lane}, B.~F. and {Muterspaugh}, M.~W., ``{Differential Astrometry of
  Subarcsecond Scale Binaries at the Palomar Testbed Interferometer},'' {\em
  \apj}~{\bf 601},  1129--1135 (2004).

\bibitem{Muterspaugh:2010lr2}
{Muterspaugh}, M.~W., {Lane}, B.~F., {Kulkarni}, S.~R., et~al., ``{The Phases
  Differential Astrometry Data Archive. V. Candidate Substellar Companions to
  Binary Systems},'' {\em \aj}~{\bf 140},  1657--1671 (2010).

\bibitem{Kok:2013uq}
{Kok}, Y., {Maestro}, V., {Ireland}, M.~J., et~al., ``{Simulating a dual beam
  combiner at SUSI for narrow-angle astrometry},'' {\em Experimental Astronomy}
   (2013).

\bibitem{Fomalont:2003ys}
{Fomalont}, E.~B. and {Kopeikin}, S.~M., ``{The Measurement of the Light
  Deflection from Jupiter: Experimental Results},'' {\em \apj}~{\bf 598},
  704--711 (2003).

\bibitem{Reid:2004ve}
{Reid}, M.~J. and {Brunthaler}, A., ``{The Proper Motion of Sagittarius A*. II.
  The Mass of Sagittarius A*},'' {\em \apj}~{\bf 616},  872--884 (2004).

\bibitem{Reid:2009uq}
{Reid}, M.~J., {Menten}, K.~M., {Brunthaler}, A., et~al., ``{Trigonometric
  Parallaxes of Massive Star-Forming Regions. I. S 252 \& G232.6+1.0},'' {\em
  \apj}~{\bf 693},  397--405 (2009).

\bibitem{Bower:2011fj}
{Bower}, G.~C., {Bolatto}, A., {Ford}, E.~B., et~al., ``{Radio Interferometric
  Planet Search. II. Constraints on Sub-jupiter-mass Companions to GJ 896A},''
  {\em \apj}~{\bf 740},  32 (2011).

\bibitem{Forbrich:2013uq}
{Forbrich}, J., {Berger}, E., and {Reid}, M.~J., ``{An Astrometric Search for a
  Sub-stellar Companion of the M8.5 Dwarf TVLM 513-46546 Using Very Long
  Baseline Interferometry},'' {\em ArXiv e-prints}  (2013).

\bibitem{Launhardt2008}
{Launhardt}, R., {Queloz}, D., {Henning}, T., et~al., ``{The ESPRI project:
  astrometric exoplanet search with PRIMA},'' in [{\em Proc. SPIE
  7013}{\nolinebreak\hspace{0.1em}]},  (2008).

\bibitem{Sahlmann:2012uq}
{Sahlmann}, J., {S{\'e}gransan}, D., {M{\'e}rand}, A., et~al., ``{Narrow-angle
  astrometry with PRIMA},'' in [{\em Proc. SPIE
  8445}{\nolinebreak\hspace{0.1em}]},  (2012).

\bibitem{Appenzeller:1998lr}
{Appenzeller}, I., {Fricke}, K., {F{\"u}rtig}, W., et~al., ``{Successful
  commissioning of FORS1 - the first optical instrument on the VLT.},'' {\em
  The Messenger}~{\bf 94},  1--6 (1998).

\bibitem{Lazorenko:2006qf}
{Lazorenko}, P.~F., ``{Astrometric precision of observations at VLT/FORS2},''
  {\em \aap}~{\bf 449},  1271--1279 (2006).

\bibitem{Lazorenko:2007ul}
{Lazorenko}, P.~F., {Mayor}, M., {Dominik}, M., et~al., ``{High-precision
  astrometry on the VLT/FORS1 at time scales of few days},'' {\em \aap}~{\bf
  471},  1057--1067 (2007).

\bibitem{Lazorenko:2011lr}
{Lazorenko}, P.~F., {Sahlmann}, J., {S\'egransan}, D., et~al., ``{Astrometric
  search for a planet around VB 10},'' {\em \aap}~{\bf 527},  A25+ (2011).

\bibitem{Sahlmann:2013kk}
{Sahlmann}, J., {Lazorenko}, P.~F., {Segransan}, D., et~al., ``{Astrometric
  orbit of a low-mass companion to an ultracool dwarf},'' {\em ArXiv e-prints}
  (2013).

\bibitem{Guyon:2012fk}
{Guyon}, O., {Bendek}, E.~A., {Eisner}, J.~A., et~al., ``{High-precision
  Astrometry with a Diffractive Pupil Telescope},'' {\em \apjs}~{\bf 200},  11
  (2012).

\bibitem{Perryman:2001vn}
{Perryman}, M.~A.~C., {de Boer}, K.~S., {Gilmore}, G., et~al., ``{GAIA:
  Composition, formation and evolution of the Galaxy},'' {\em \aap}~{\bf 369},
  339--363 (2001).

\bibitem{de-Bruijne:2012kx}
{de Bruijne}, J.~H.~J., ``{Science performance of Gaia, ESA's space-astrometry
  mission},'' {\em \apss}~{\bf 341},  31--41 (2012).

\bibitem{2011AdSpR..47..356M}
{Mignard}, F., ``{A few metrological aspects of the Gaia mission},'' {\em
  Advances in Space Research}~{\bf 47},  356--364 (2011).

\bibitem{Casertano:2008th}
{Casertano}, S., {Lattanzi}, M.~G., {Sozzetti}, A., et~al., ``{Double-blind
  test program for astrometric planet detection with Gaia},'' {\em \aap}~{\bf
  482},  699--729 (2008).

\bibitem{Meyer:2011fj}
{Meyer}, E., {K{\"u}rster}, M., {Arcidiacono}, C., et~al., ``{Astrometry with
  the MCAO instrument MAD. An analysis of single-epoch data obtained in the
  layer-oriented mode},'' {\em \aap}~{\bf 532},  A16 (2011).

\bibitem{Rigaut:2012zr}
{Rigaut}, F., {Neichel}, B., {Boccas}, M., et~al., ``{GeMS: first on-sky
  results},'' in [{\em Proc. SPIE 8847}{\nolinebreak\hspace{0.1em}]},  (2012).

\bibitem{Paufique:2012kx}
{Paufique}, J., {Argomedo}, J., {Arsenault}, R., et~al., ``{Status of the GRAAL
  system development: very wide-field correction with 4 laser guide-stars},''
  in [{\em Proc. SPIE 8847}{\nolinebreak\hspace{0.1em}]},  (2012).

\bibitem{Trippe:2010fj}
{Trippe}, S., {Davies}, R., {Eisenhauer}, F., et~al., ``{High-precision
  astrometry with MICADO at the European Extremely Large Telescope},'' {\em
  \mnras}~{\bf 402},  1126--1140 (2010).

\bibitem{Unwin:2008ys}
{Unwin}, S.~C., {Shao}, M., {Tanner}, A.~M., et~al., ``{Taking the Measure of
  the Universe: Precision Astrometry with SIM PlanetQuest},'' {\em \pasp}~{\bf
  120},  38--88 (2008).

\end{thebibliography}
\bibliographystyle{spiebib_mod}   
\end{document}